\documentstyle[aps,prb,multicol,psfig]{revtex}
\begin{document}
\title{
Spin transport properties of the quantum one-dimensional
non-linear sigma model
: an application to Haldane gap systems
}
\author{Satoshi Fujimoto}
\address{
Department of Physics,
Kyoto University, Kyoto 606, Japan
}
\date{\today}
\maketitle

\begin{abstract}
Spin transport properties of the quantum one-dimensional non-linear
sigma model are studied based upon the Bethe ansatz exact solution
for the $O(3)$ sigma model and the $1/N$-expansion approach 
for the $O(N)$ sigma model.
It is shown that the spin transport at finite temperatures is
ballistic and there is no spin diffusion in this system.
A possible mechanism for the spin diffusion observed in the NMR
experiment of the Haldane gap system ${\rm AgVP_2S_6}$ 
by Takigawa {\it et al.} [Phys. Rev. Lett.{\bf 76}, 2173 (1996)]
is also discussed.
The inclusion of the spin-phonon interaction accounts for the experimental 
results qualitatively.
\end{abstract}



\section{Introduction}

The spin diffusion in the Heisenberg spin chain is a long-standing issue
in the studies of magnetism.\cite{vis}
For the XXZ Heisenberg chain with spin $s=1/2$, there is 
a strong piece of evidence for the absence of the spin diffusion 
in the massless state.\cite{mcl,zotos,naro,mccoy}  
In the case that the spin $s$ is an integer, 
there exists an excitation gap, as predicted 
by Haldane many years ago using the mapping to
the one-dimensional(1D) quantum $O(3)$ non-linear sigma model.\cite{hal}
Thus it is expected that the spin transport properties at low temperatures
are quite different from those of the $s=1/2$ Heisenberg chain. 
Recently, it was observed in the NMR experiment for
the Haldane gap system, ${\rm AgVP_2S_6}$, by Takigawa {\it et al.} 
that a diffusive behavior appears in long wavelength 
spin-spin correlation functions.\cite{taki}
In order to explain the experiment,
Sachdev and Damle showed that a semiclassical analysis of
the quantum 1D $O(3)$ non-linear sigma model gives
a diffusive behavior in the spin correlation function 
at low temperatures.\cite{sac}
However, the 1D quantum $O(3)$ sigma model is an integrable model,
and exactly solvable in terms of the Bethe ansatz method.\cite{wie}
The spin current is conserved, and thus it is expected that
the spin diffusion is absent in this model.
Although we can not calculate correlation functions directly
by using the Bethe ansatz method,
the spin transport properties are also probed by the Drude weight,
namely, the zero frequency part of the spin conductivity.
The calculation of the Drude weight for exactly solvable models
was recently extended to the finite temperature case 
by Fujimoto and Kawakami.\cite{sf}
If the Drude weight is non-zero at finite temperatures, 
the system shows a ballistic transport rather than a diffusive one.
One of the purposes of this paper is to discuss 
the exact Drude weight of the 1D quantum $O(3)$ sigma model 
at finite temperatures based upon the Bethe ansatz method.
Our finding is that the Drude weight {\it is non-vanishing}
at finite temperatures, and that the spin transport is ballistic.
We will also confirm this result by utilizing the $1/N$-expansion
analysis for the $O(N)$ non-linear sigma model.
For large $N$, we can calculate the spin correlation function
perturbatively.
We will show that the Ward-Takahashi identity 
for the current vertex operators
leads to a ballistic form of the spin correlation function, and that
the spin diffusion does not exist.
Thus in order to explain the spin diffusion observed in 
the Haldane gap system ${\rm AgVP_2S_6}$,
we need to introduce a coupling with some other degrees of freedom,
such as interchain couplings, or spin-phonon interaction, etc.
Moreover a remarkable feature of this experiment is the existence
of a temperature-dependent cutoff frequency, which can not be
explained by the pure non-linear sigma model.  
The second purpose of this paper is to investigate 
a possible mechanism of the spin diffusion in this system.
The interchain coupling $J'$ of this material is so small, 
$J'/J\leq 10^{-5}$, that it is unlikely to give rise 
to the observed spin diffusion.\cite{taki}
We will try to explain the experimental results by introducing
the coupling with phonons.
We will show that the inclusion of the spin-phonon interaction 
gives, at least qualitatively, a fine explanation for the spin diffusion
and the presence of the temperature-dependent cutoff 
frequency observed in ${\rm AgVP_2S_6}$.

The organization of this paper is as follows.
In Sec.II, the exact Drude weight at finite temperatures 
is derived based upon the Bethe ansatz solution.
In Sec.III, the non-diffusive behavior of the spin 
correlation function is argued by using the $1/N$-expansion 
of the $O(N)$ sigma model.
In Sec.IV, we discuss a possible mechanism for the spin diffusion 
observed in the NMR experiment taking into account a coupling with phonons.
Summary and discussions are given in Sec.V. 

\section{The Bethe Ansatz Exact Solution for the
$O(3)$ Sigma Model and the Finite Temperature Drude Weight}

In this section, we will derive the exact Drude weight of
the quantum 1D $O(3)$  non-linear sigma model at finite temperatures
using the Bethe ansatz exact solution.
The action of the $O(3)$ non-linear sigma model is given by,
\begin{equation}
S=\frac{1}{2g}\int{\rm d}x {\rm d}\tau v(\partial_{\mu}\mbox{\boldmath $n$})^2,
\end{equation} 
where $\mbox{\boldmath $n$}=(n_x,n_y,n_z)$ is a unit vector, 
$(\mbox{\boldmath $n$})^2=1$.
The elementary excitations of this model
are massive triplets interacting with each other.\cite{zam}
The Bethe ansatz exact solution for this model 
was obtained by Wiegmann a decade ago.\cite{wie}
In order to obtain the Drude weight for the spin conductivity,
we need to use the twisted boundary condition introducing 
the flux $\Phi$ which couples with spin degrees of 
freedom.\cite{shastry,by} 
The Bethe ansatz equations under the twisted boundary conditions 
are given by,\cite{wie}
\begin{eqnarray}
e^{{\rm i}m\sinh \theta_{\alpha}L}&=e^{{\rm i}\Phi}&\prod^p_{\beta=1}
\frac{\theta_{\alpha}-\theta_{\beta}-{\rm i}\pi}
{\theta_{\alpha}-\theta_{\beta}+{\rm i}\pi} 
\prod^q_{\gamma=1}\frac{\theta_{\alpha}-\lambda_{\gamma}+{\rm i}\pi}
{\theta_{\alpha}-\lambda_{\gamma}-{\rm i}\pi}, 
\label{bae1} \\
\prod^p_{\beta=1}\frac{\theta_{\beta}-\lambda_{\alpha}+{\rm i}\pi}
{\theta_{\beta}-\lambda_{\alpha}-{\rm i}\pi}
&=&-e^{-2{\rm i}\Phi}\prod^q_{\gamma=1}
\frac{\lambda_{\alpha}-\lambda_{\beta}+{\rm i}\pi}
{\lambda_{\alpha}-\lambda_{\beta}-{\rm i}\pi},
\label{bae2}
\end{eqnarray}
where $p$ and $q$ are defined by the eigen values of 
the Noether current operator 
$\mbox{\boldmath $l$}=\mbox{\boldmath $n$}
\times\partial_t\mbox{\boldmath $n$}/v$;
$\mbox{\boldmath $l$}^2=p(p+1)$, $l_z=p-q$, and
$m\propto \exp(-2\pi/g)/g$. 
The total energy in the presence of a magnetic field $h$ is given by
\begin{equation}
E=m\sum_{\alpha}\cosh\theta_{\alpha}-h(p-q).
\label{energy} 
\end{equation}
The thermodynamic properties are obtained by using the
string hypothesis for the solutions of the rapidities.
As was shown by Wiegmann, $\theta_{\alpha}$ is real, and
$\lambda^{n,k}_{\alpha}=\lambda^n_{\alpha}+{\rm i}\pi(n+1-2k)/2$, 
$k=1,2,...,n$.\cite{wie}
The Drude weight at finite temperatures is derived from
the finite size corrections to the thermodynamic Bethe ansatz equations.
In order to calculate the finite size corrections, 
we expand the real part of the rapidities in $1/L$,\cite{sf,bm}
\begin{eqnarray}
\theta_{\alpha}&=&\theta_{0 \alpha}+\frac{f_1}{L}+\frac{f_2}{L^2}
+O(1/L^3),  \\
\lambda^n_{\alpha}&=&\lambda^{n}_{0 \alpha}
+\frac{g_{1n}}{L}+\frac{g_{2n}}{L^2}+O(1/L^3).
\end{eqnarray}
Substituting these rapidities into eqs.(\ref{bae1}) and (\ref{bae2}), and 
taking a continuum limit, we obtain the coupled integral equations 
which determine the spectrum of the system.
The zeroth order contributions in $1/L$ give the conventional 
thermodynamic Bethe ansatz equations which were 
first obtained by Tsvelik,\cite{tsv}
\begin{eqnarray}
\rho(\theta)+\rho^h(\theta)&=&\frac{m}{2\pi}\cosh\theta
-\int^{\infty}_{-\infty}{\rm d}\theta'\frac{1}{(\theta-\theta')^2+\pi^2}
\rho(\theta') \nonumber \\
&&+\sum_{n=1}^{\infty}\int^{\infty}_{-\infty}{\rm d}\lambda
a_n(\theta-\lambda)\sigma_n(\lambda), 
\label{dis1} \\
\sigma^h_n(\lambda)+\sum_{m=1}^{\infty}A_{nm}*\sigma_m(\lambda)
&=&\int^{\infty}_{-\infty}{\rm d}\theta a_n(\theta-\lambda)\rho(\theta), 
\label{dis2}\\
\kappa(\theta)&=&m\cosh\theta-h-T\int^{\infty}_{-\infty} {\rm d}\theta'
\frac{1}{(\theta-\theta')^2+\pi^2}{\rm ln}(1+e^{-\frac{\kappa(\theta')}{T}}) 
\nonumber \\
&-&T\sum_{n=1}^{\infty}\int^{\infty}_{-\infty}{\rm d}
\lambda a_n(\theta-\lambda)
{\rm ln}(1+e^{-\frac{\varepsilon_n(\lambda)}{T}}), \label{tba1} \\
{\rm ln}(1+e^{\frac{\varepsilon_n(\lambda)}{T}})
&=&\frac{nh}{T}-\int^{\infty}_{-\infty}{\rm d}\theta a_n(\theta-\lambda)
{\rm ln}(1+e^{-\frac{\kappa(\theta)}{T}}) \nonumber \\
&+&\sum_{m}A_{nm}*{\rm ln}(1+e^{-\frac{\varepsilon_n(\lambda)}{T}}).
\label{tba2}
\end{eqnarray}
Here $\rho(\theta)$ and $\sigma_n(\lambda)$ are the distribution functions
for the rapidities $\theta_{0 \alpha}$ and $\lambda^{n}_{0 \alpha}$,
respectively, and $\rho^h(\theta)$ and $\sigma^h_n(\lambda)$
are those for holes.
$\kappa(\theta)\equiv T{\rm ln}(\rho^h/\rho)$,
$\varepsilon_n(\lambda)\equiv T{\rm ln}(\sigma^h_n/\sigma_n)$,
and
$$a_n(x)=\frac{2(n-1)}{4x^2+\pi^2(n-1)^2}
+\frac{2(n+1)}{4x^2+\pi^2(n+1)^2},$$
$$A_{nm}*\phi(x)=\delta_{nm}\phi(x)+\frac{\rm d}{{\rm d}x}
\int^{\infty}_{-\infty}\frac{{\rm d}x'}{2\pi}
\Theta_{nm}(\frac{x-x'}{\pi/2})\phi(x'),$$
with 
\begin{equation}
\Theta_{nm}(x)=\left\{
\begin{array}{ll}
\displaystyle{\theta\biggl(\frac{x}{\vert n-m\vert}\biggr)
+2\theta\biggl(\frac{x}{\vert n-m\vert +2}\biggr)
+\cdot\cdot\cdot
+2\theta\biggl(\frac{x}{ n+m-2}\biggr)
+\theta\biggl(\frac{x}{n+m}\biggr)}&\qquad n\neq m \\
\displaystyle{2\theta\biggl(\frac{x}{2}\biggr)+\cdot\cdot\cdot
+2\theta\biggl(
\frac{x}{2n-2}\biggr)+\theta\biggl(\frac{x}{2n}\biggr)}&\qquad n=m,
\end{array}
\right.
\end{equation}
and $\theta(x)=2\tan^{-1}x$.

The finite size corrections to the above equations give
the coupled equations which determine $f_{1,2}$ and $g_{1,2n}$,
\begin{eqnarray}
&&(\rho(\theta)+\rho^h(\theta))f_1(\theta)=\frac{\Phi}{2\pi}
-\int^{\infty}_{-\infty}{\rm d}\theta'\frac{1}{(\theta-\theta')^2+\pi^2}
\rho(\theta')f_1(\theta') \nonumber \\
&&+\sum_{n=1}^{\infty}\int^{\infty}_{-\infty}{\rm d}\lambda
a_n(\theta-\lambda)\sigma_n(\lambda)g_{1n}(\lambda), \label{fsc1} \\
&&\sigma^h_n(\lambda)g_{1n}(\lambda)+\sum_{m=1}^{\infty}
A_{nm}*\sigma_m(\lambda)g_{1m}(\lambda)
=\int^{\infty}_{-\infty}{\rm d}\theta
a_n(\theta-\lambda)\rho(\theta)f_1(\theta)-\frac{n\Phi}{\pi}, \label{fsc2} \\
&&(\rho(\theta)+\rho^h(\theta))f_2(\theta)=
-\int^{\infty}_{-\infty}{\rm d}\theta'\frac{1}{(\theta-\theta')^2+\pi^2}
\rho(\theta')f_2(\theta') \nonumber \\
&&+\sum_{n=1}^{\infty}\int^{\infty}_{-\infty}{\rm d}\lambda
a_n(\theta-\lambda)\sigma_n(\lambda)g_{2n}(\lambda) 
+\frac{1}{2}\frac{\rm d}{{\rm d}\theta}
[(\rho(\theta)+\rho^h(\theta))f^2_1(\theta)]
\nonumber \\
&&-\int^{\infty}_{-\infty}{\rm d}\theta'\frac{\theta-\theta'}
{((\theta-\theta')^2+\pi^2)^2}\rho(\theta')f^2_1(\theta')
+\sum_{n=1}^{\infty}\int^{\infty}_{-\infty}{\rm d}\lambda
\frac{{\rm d} a_n(\theta-\lambda)}{{\rm d}\lambda}
\sigma_n(\lambda)g^2_{1n}(\lambda) \nonumber \\
&&+(\mbox{surface terms}), \label{fsc3}\\
&&\sigma^h_n(\lambda)g_{2n}(\lambda)+\sum_{m=1}^{\infty}A_{nm}
*\sigma_m(\lambda)g_{2m}(\lambda)
=\int^{\infty}_{-\infty}{\rm d}\theta
a_n(\theta-\lambda)\rho(\theta)f_2(\theta) \nonumber \\
&&+\frac{1}{2}\frac{{\rm d}}{{\rm d}\theta}
[(\sigma(\lambda)+\sigma^h(\lambda))g^2_{1n}]
-\sum_m \frac{{\rm d} A_{nm}}{{\rm d} \lambda}*\sigma_m g^2_{1m}
+\int^{\infty}_{-\infty}{\rm d}\theta\frac{{\rm d} 
a_n(\theta-\lambda)}{{\rm d}\theta}
\rho(\theta) f^2_1(\theta) \nonumber \\
&&+(\mbox{surface terms}). \label{fsc4}
\end{eqnarray}

The Drude weight at finite temperatures is given by the second 
derivative of the energy spectrum  with respect 
to the Aharonov-Bohm flux $\Phi$,\cite{kohn}
\begin{equation}
D=\frac{L}{2}\biggl\langle\frac{{\rm d}^2 E_n}{ {\rm d} \Phi^2}
\biggr\rangle\biggl\vert_{\Phi=0}\biggl..\label{dru}
\end{equation}  
Here $\langle \cdot\cdot\cdot\rangle$ is a thermal average.
Note that the Drude weight $D$ defined above is different from 
the second derivative of 
the free energy which is related to a Meissner fraction.\cite{gs}

Evaluating the finite size corrections to 
the total energy up to the second order in $1/L$,\cite{we}
\begin{equation}
\frac{E}{L}=E_0+\frac{E_1}{L}
+\frac{E_2}{L^2},
\end{equation}
we find that the dependence on $\Phi$ appears in the second order term
$E_2$.
Then from eqs.(\ref{energy}) and (\ref{dis1})-(\ref{fsc4}),
we obtain the formula for the Drude weight,
\begin{eqnarray}
D&=&\frac{1}{2}\frac{{\rm d}^2E_2}{{\rm d}\Phi^2}
\biggl\vert_{\Phi=0}\biggr. \nonumber \\
&=&\frac{1}{2}\int^{\infty}_{-\infty} {\rm d}\theta
\biggl\{(\rho(\theta)+\rho^h(\theta))
\frac{{\rm d} f_1}{ {\rm d}\Phi}\biggr\}^2
\frac{ {\rm d}}{{\rm d}\theta}\biggl(
\frac{-1}{1+e^{\kappa(\theta)/T}}\biggr)
\frac{1}{\rho(\theta)+\rho^h(\theta)}
\frac{{\rm d}\kappa(\theta)}{{\rm d}\theta} 
\nonumber \\
&+&\frac{1}{2}\sum_{n=1}^{\infty}\int^{\infty}_{-\infty}
{\rm d}\lambda
\biggl\{(\sigma_n(\lambda)+\sigma_n^h(\lambda))
\frac{{\rm d} g_{1n}}{{\rm d} \Phi}
\biggr\}^2\frac{{\rm d}}{{\rm d}\lambda}
\biggl(\frac{-1}{1+e^{\varepsilon_n(\lambda)/T}}\biggr)
\frac{1}{\sigma_n(\lambda)+\sigma^h_n(\lambda)}
\frac{{\rm d}\varepsilon_n(\lambda)}{{\rm d}\lambda}.
\label{drude}
\end{eqnarray}
This expression is similar to that for the Hubbard model.\cite{sf}

\subsection{Low temperature expansion}

We now derive the temperature dependence of $D$ 
in the low temperature limit.
We consider the case of the non-vanishing magnetic field
$h\neq 0$.
In this case, from eq.(\ref{tba2}) we find $\varepsilon_n(\lambda)>0$.
Thus, at low temperatures, eq.(\ref{tba1}) is reduced to 
\begin{equation}
\kappa(\theta)=m\cosh\theta-h+\int^{B}_{-B}{\rm d}\theta'
\frac{1}{(\theta-\theta')^2+\pi^2}\kappa(\theta'),
\end{equation}
where $\kappa(\pm B)=0$.
In the case of $h<m$, all excitations are massive, and
the solution for the above equation is given by
\begin{equation}
\kappa(\theta)=m\cosh\theta-h,
\end{equation}
and $B=0$.
The low temperature limit of eq.(\ref{tba2}) gives
$\varepsilon_n(\lambda)=nh$ and 
${\rm d}\varepsilon_n(\lambda)/{\rm d}\lambda\sim \exp(-(m-h)/T)$. From
eqs.(\ref{dis1}) and (\ref{dis2}), we have
\begin{eqnarray}
\rho(\theta)+\rho^h(\theta)&=&\frac{m}{2\pi}\cosh\theta, \\
\sigma_n(\lambda)+\sigma_n^h(\lambda)&=&e^{-\frac{m-h}{T}}. \label{dislow}
\end{eqnarray}
Moreover using eqs.(\ref{fsc1}) and (\ref{fsc2}),
we can show that at sufficiently low temperatures 
\begin{eqnarray}
(\rho(\theta)+\rho^h(\theta))\frac{{\rm d} f_1(\theta)}{{\rm d} \Phi}&\sim&
\frac{1}{2\pi}, \\
(\sigma_n(\lambda)+\sigma_n^h(\lambda))\frac{{\rm d} g_{1n}}{{\rm d} \Phi}
&\sim&-\frac{n}{\pi}.
\end{eqnarray} 
Substituting these expressions into eq.(\ref{drude}),
we obtain the Drude weight in the low temperature limit,
\begin{equation}
D\sim \sqrt{T}e^{-\frac{m-h}{T}}.
\end{equation}
Thus the Drude weight is non-zero at finite temperatures.
This result implies that the spin transport of the 1D quantum $O(3)$
non-linear sigma model in the low temperature region is
ballistic, and that the spin diffusion does not exist.

\subsection{High temperature expansion}

Here we consider the Drude weight 
in the high temperature limit.
Although in the sufficiently high temperature region, 
the non-linear sigma model is not related to
the $s=1$ Heisenberg spin chain,
it is interesting to check that
the spin diffusion does not occur in the non-linear
sigma model even in the high temperature limit.
We invoke the standard method for the high temperature expansion
of the Bethe ansatz solution developed by Takahashi.\cite{taka}
For this purpose, it is convenient to
rewrite eqs.(\ref{dis1}), (\ref{dis2}), (\ref{tba1}) and 
(\ref{tba2}) into the following form.
\begin{eqnarray}
\rho+\rho^h&=&\frac{m}{2\pi}\cosh\theta-s*\sigma_2^h, \label{tdis1}\\
\sigma_n+\sigma_n^h&=&\delta_{2n}s*\rho+
s*(\sigma_{n-1}^h+\sigma_{n+1}^h),
 \qquad n\geq 2, \label{tdis2}\\
\sigma_1+\sigma_1^h&=&s*\sigma_2, \label{tdis3} \\
\kappa(\theta)&=&m\cosh\theta-Ts*{\rm ln}
(1+e^{\frac{\varepsilon_2(\lambda)}{T}}), \label{ttba1}\\
\varepsilon_n(\lambda)&=&Ts*{\rm ln}
(1+e^{\frac{\varepsilon_{n-1}(\lambda)}{T}})
(1+e^{\frac{\varepsilon_{n+1}(\lambda)}{T}}) \nonumber \\
&&-\delta_{n,2}Ts*{\rm ln}(1+e^{-\frac{\kappa(\theta)}{T}}), 
\quad n=1,2,... \\
&& \lim_{n\rightarrow \infty}\frac{\varepsilon_n}{n}=h,
\label{ttba2}
\end{eqnarray}
where $\varepsilon_0\equiv -\infty$, and 
$$s*f(x)=\int^{\infty}_{-\infty}{\rm d}x'\frac{1}{4\cosh[\pi(x-x')/2]}f(x').$$
Since the non-linear sigma model is a relativistic field theory
model, it is appropriate to introduce the ultra-violet(UV) cutoff $\Lambda$
which may be of order $J$, the exchange energy of 
the original $s=1$ Heisenberg spin chain.
In the case of $T\gg\Lambda$, eqs.(\ref{ttba1}) and (\ref{ttba2}) are
recasted into,
\begin{eqnarray}
\ln \zeta&=&-\frac{1}{2}\ln (1+\eta_2), \\
\ln\eta_2&=&\frac{1}{2}\ln(1+\eta_1)(1+\eta_3)
-\frac{1}{2}\ln(1+\zeta^{-1}), \\
\ln\eta_n&=&\frac{1}{2}\ln(1+\eta_{n-1})(1+\eta_{n+1}), \qquad n \geq 3, \\
&&\lim_{n\rightarrow\infty}\frac{\ln\eta_n}{n}\rightarrow 0,
\end{eqnarray}
where $\zeta=\rho^h/\rho$, $\eta_n=\sigma^h_n/\sigma_n$.
Thus in this limit, $\zeta$ and $\eta_n$ are constants, and 
the solutions of the above equations are given by
\begin{eqnarray}
\eta_n&=&n^2-1 \qquad n\geq 2, \nonumber \\
\eta_1&=&2, \qquad \zeta=\frac{1}{2}. \\
\end{eqnarray}
Substituting these solutions into eqs.(\ref{tdis1}), (\ref{tdis2}),
and (\ref{tdis3}), we can solve these coupled equations for 
$\rho$ and $\sigma_n$ by using the standard method.
Then we find,
\begin{eqnarray}
\sigma_n(\lambda)&=&\frac{m}{6\pi n}
\int^{\theta_c}_{-\theta_c} 
\frac{{\rm d}\theta}{2\pi}\biggl[\frac{1}{(\theta-\lambda)^2+(n-1)^2}
-\frac{1}{(\theta-\lambda)^2+(n+1)^2}\biggr]\cosh\theta, 
\qquad n\geq 3,  \label{hsig}\\
\sigma_2(\lambda)&=&\frac{m}{12\pi}s*\cosh\theta+2s*\sigma_3(\lambda), \\
\sigma_1(\lambda)&=&s*\sigma_2(\lambda), \\
\rho(\theta)&=&\frac{m}{3\pi}-2s*\sigma_2(\lambda).
\end{eqnarray}
Here we have introduced the UV cutoff $\theta_c\equiv \cosh^{-1}(\Lambda/m)$
in order to suppress the divergence of the $\theta$-integral
in eq.(\ref{hsig}).
We can also obtain $ {\rm d} f_1/ {\rm d} \Phi$ and 
$ {\rm d} g_{1n}/ {\rm d} \Phi$ 
from eqs.(\ref{fsc1}) and (\ref{fsc2}) in a similar manner.
The results are 
\begin{eqnarray}
\rho  \frac{{\rm d} f_1}{ {\rm d}\Phi}&=&-\frac{5}{19\pi}, \qquad
\sigma_1\frac{ {\rm d} g_{11}}{ {\rm d}\Phi}=-\frac{1}{36\pi}, \qquad
\sigma_2\frac{ {\rm d} g_{12}}{ {\rm d}\Phi}=-\frac{1}{18\pi}, \\
\sigma_n\frac{ {\rm d} g_{1n}}{ {\rm d}\Phi}&=&-\frac{1}{3\pi n(n^2-1)}, 
\qquad n\geq 3.
\end{eqnarray}
Furthermore in the high temperature limit,
we have the following relations,
\begin{eqnarray}
\rho(\theta)&=&\frac{1}{2\pi}{\rm sgn}\theta
\frac{{\rm d}\kappa}{{\rm d}\theta}
\frac{1}{1+e^{\frac{\kappa}{T}}}, \\
\rho^h(\theta)&=&\frac{1}{2\pi}{\rm sgn}\theta
\frac{{\rm d}\kappa}{{\rm d}\theta}
\frac{e^{\frac{\kappa}{T}}}{1+e^{\frac{\kappa}{T}}}, \\
\sigma_n(\lambda)&=&\frac{1}{2\pi}{\rm sgn}\lambda
\frac{{\rm d}\varepsilon_n}{{\rm d}\lambda}
\frac{1}{1+e^{\frac{\varepsilon_n}{T}}}, \\
\sigma_n^h(\lambda)&=&\frac{1}{2\pi}{\rm sgn}\lambda
\frac{{\rm d}\varepsilon_n}{{\rm d}\lambda}
\frac{e^{\frac{\varepsilon_n}{T}}}{1+e^{\frac{\varepsilon_n}{T}}}.
\end{eqnarray}
Then substituting these equations into eq.(\ref{drude}), 
we have the Drude weight,
\begin{equation}
D\sim \frac{C}{T},
\end{equation}
with 
\begin{equation}
C=\frac{25}{486}\int {\rm d}\theta\rho(\theta)+
\frac{1}{972}\int {\rm d}\lambda \sigma_1(\lambda)
+\frac{3}{648}\int {\rm d}\lambda\sigma_2(\lambda)
+\sum_{n=3}^{\infty}\frac{2}{9n^4(n^2-1)}\int {\rm d}\lambda\sigma_n(\lambda).
\end{equation}
Note that $D$ in the high temperature limit has
the same temperature dependence as that of the 1D Hubbard model.
Since even in the high temperature limit, the Drude weight is non-zero,
it is expected to be non-vanishing in the whole temperature region.
Therefore the spin diffusion does not exist in the 1D quantum
$O(3)$ non-linear sigma model.
This is the main conclusion of this section.

\section{$1/N$-Expansion Approach to the Spin Dynamics of
the $O(N)$ Sigma Model} 

In the previous section, we have calculated
the Drude weight at finite temperatures which is a probe for transport
properties.
However in order to discuss the spin diffusion,
it is desirable to study spin-spin correlation functions directly.
This issue is not tractable in terms of the Bethe ansatz method.
Moreover, there is a criticism by Buragohain and Sachdev  
against the approach exploited in the previous section.\cite{sac2}
They claimed that the calculation of the Drude weight
from the finite size spectrum may not give the correct
thermodynamic limit at finite temperatures.
Thus, in this section, we will discuss spin-spin correlation functions
with the use of the $1/N$ expansion method for
the quantum 1D $O(N)$ sigma model,\cite{pol,bre}
and check the consistency with the results obtained by the Bethe ansatz
method in Sec.II.
The $1/N$-expansion is utilized by several authors in the study
of Haldane gap systems.\cite{aff,jol,sagi}
Although the $O(N)$ sigma model with large $N$ 
is not related to the Haldane gap systems,
the qualitative properties of the excitation are similar 
to those of the $O(3)$ sigma model.
It is expected that the results obtained for the $O(N)$ sigma model
with large $N$ may give a qualitative understanding of the $O(3)$
sigma model.

We are mainly concerned with the spin lattice relaxation rate $1/T_1$,
which is given by
\begin{equation}
\frac{1}{T_1T}\propto\lim_{\omega\rightarrow 0}\sum_q A^2(q)
\frac{{\rm Im}\chi(q,\omega)}{\omega},
\end{equation}
where $\chi(q,\omega)$ is the retarded spin correlation function,
$\langle S^z(q,\omega)S^z(-q,-\omega)\rangle^R$,
and $A(q)$ is a structure factor.
It is known that in Haldane gap systems, the main contribution to
$1/T_1$ comes from the uniform part of
spin-spin correlation functions for $q\sim 0$.\cite{taki,sagi}
It is given by the correlation function
of the uniform magnetization vector 
$\mbox{\boldmath $l$}=\mbox{\boldmath $n$}
\times\partial_t\mbox{\boldmath $n$}/v$.
That is, 
\begin{equation}
\chi(q\sim 0,\omega)=
\int^{\infty}_{-\infty} { d}x\int^{\infty}_0{ d}t 
\langle [l^{\alpha}(x,t),l^{\alpha}(0,0)]\rangle
e^{{\rm i}(\omega t+qx)}.
\label{spinc}
\end{equation}

For sufficiently large $N$, the excitation
in the $O(N)$ sigma model is massive $N$-plet bosons 
interacting weakly with each other via a coupling constant of order
$O(1/N)$.\cite{pol}
Thus we can calculate the correlation function 
by the perturbative expansion in terms of 
$1/N$.
The Green's function of the free massive $N$-plet bosons 
$\mbox{\boldmath $n$}$
is given by,
\begin{equation}
G^{\alpha}_0(p,\omega_n)\equiv \int^{\beta}_{0}{\rm d}\tau
\langle T_{\tau} n^{\alpha}(q,\tau)n^{\alpha}(-q,0)\rangle 
e^{{\rm i}\omega_n \tau}
=\frac{1}{({\rm i}\omega_n)^2-E_p^2},
\label{magnon}
\end{equation}
where $\omega_n=2\pi nT$, and $E_p=\sqrt{v^2p^2+m^2}$.
One of the important effects of the interaction is
the quasiparticle damping which is given by the imaginary part of
the self-energy.
The lowest order self-energy correction of which 
the diagram is shown in Fig.1, is easily evaluated as
${\rm Im}\Sigma^R \sim c(\omega/T)\exp(-2m/T)$ ($c$, a constant) 
for small $\omega$
and $p=0$.\cite{jol} 
If this damping effect is sufficiently large, 
it leads to the diffusive form
of the correlator for the $\vec{n}$ field, 
\begin{equation}
G^{\alpha R}(p,\omega)
=\frac{1}{(\omega+{\rm i}\delta)^2-v^2p^2-m^2-{\rm i}c(\omega/T)\exp(-2m/T)}.
\end{equation}
Naively thinking, it seems that
this correlation function gives rise to a diffusive
behavior of $1/T_1$.
However this expectation is not correct.
$\mbox{\boldmath $l$}$ is the conserved current of 
the non-linear sigma model which
satisfies the continuity equation 
$\partial_t\mbox{\boldmath $l$}+\partial_x\mbox{\boldmath $j$}=0$,
where $\mbox{\boldmath $j$}\equiv v\mbox{\boldmath $n$}
\times\partial_x\mbox{\boldmath $n$}$.
Therefore a Ward-Tkahashi identity holds for these current operators.
The Ward-Takahashi identity ensures
the cancellation of the self-energy corrections with
vertex corrections of the correlation functions, which leads
to a ballistic behavior of spin-spin correlation functions.
In order to show this, we introduce the three point vertex functions
defined as,
\begin{eqnarray}
\Lambda^{\alpha}_0(p_{\mu},q_{\mu})G^{\beta}(p_{\mu})
G^{\gamma}(q_{\mu}-p_{\mu})
&=&\langle l^{\alpha}(q_{\mu})n^{\beta}(p_{\mu})
n^{\gamma}(q_{\mu}-p_{\mu})\rangle, \\
\Lambda^{\alpha}_1(p_{\mu},q_{\mu})G^{\beta}(p_{\mu})
G^{\gamma}(q_{\mu}-p_{\mu})
&=&\langle j^{\alpha}(q_{\mu})n^{\beta}(p_{\mu})
n^{\gamma}(q_{\mu}-p_{\mu})\rangle,
\end{eqnarray}
where $p_{\mu}\equiv(p,i\varepsilon_n)$ etc.
In the momentum-energy space, the current operators are expressed as
\begin{eqnarray}
l^{\alpha}(q_{\mu})&=&\sum_{p'_{\mu}}\frac{{\rm i}\varepsilon'_n}{v}
\epsilon^{\alpha\beta\gamma}n^{\beta}(q-p',\omega_m-\varepsilon'_n)
n^{\gamma}(p',\varepsilon'_n), \\
j^{\alpha}(q_{\mu})&=&\sum_{p'_{\mu}}v p'
\epsilon^{\alpha\beta\gamma}n^{\beta}(q-p',\omega_m-\varepsilon'_n)
n^{\gamma}(p',\varepsilon'_n).
\end{eqnarray}
Using the commutation relations $[l^{\alpha}(x),n^{\beta}(y)]=
{\rm i}\epsilon^{\alpha\beta\gamma}n^{\gamma}(x)\delta(x-y)$,
we can derive the Ward-Takahashi identity,\cite{wt}
\begin{equation}
{\rm i}\omega_n\Lambda_0^{\alpha}(p_{\mu},q_{\mu})+
q\Lambda_1^{\alpha}(p_{\mu},q_{\mu})
=G^{-1}(q_{\mu}-p_{\mu})
-G^{-1}(p_{\mu}). \label{wti}
\end{equation}
Here we omit the spin index $\alpha$ of $G^{\alpha}(p_{\mu})$,
since it does not depend on $\alpha$ by virtue of the $O(N)$ symmetry. 
We will take an analytic continuation 
${\rm i}\omega_n\rightarrow \omega+{\rm i}\delta$ at the last stage.
Since we are concerned with the contribution from small $q$ and $\omega$,
$\varepsilon'_n$ and $p'$ in $\Lambda^{\alpha}_{0,1}(p_{\mu},q_{\mu})$ 
can be replaced with $\varepsilon$ and $p$.
As a result, we have 
$\Lambda^{\alpha_1}(p_{\mu},q_{\mu})
\approx (v^2p/{\rm i}\varepsilon_n)\Lambda^{\alpha}_0(p_{\mu},q_{\mu})$.
Using this relation and the Ward-Takahashi identity (\ref{wti}),
we can obtain the spin-spin correlation function,
\begin{eqnarray}
\chi(q\sim 0,{\rm i}\omega_m)&=&\sum_{p_{\mu}}
\Lambda^{\alpha}_0(p_{\mu},q_{\mu})G(p_{\mu})G(q_{\mu}-p_{\mu})
{\rm i}\varepsilon_n \nonumber \\
&=&\sum_{p_{\mu}}\frac{{\rm i}\varepsilon_n}{{\rm i}\omega_m
+\frac{v^2p}{{\rm i}\varepsilon_n}q}[G(p_{\mu})-G(q_{\mu}-p_{\mu})].
\label{chi1}
\end{eqnarray}
Up to the lowest order in $1/N$, we expand $G=G_0+G_0\Sigma G_0$
in eq.(\ref{chi1}). 
Then, putting ${\rm i}\omega_m\rightarrow \omega+{\rm i}\delta$, we find that
in the limit of $q\rightarrow 0$ and $\omega\rightarrow 0$,
the above expression reduces to,
\begin{eqnarray}
\chi^R(q\sim0,\omega)&\sim& \sum_p\Bigl[
\coth\frac{E_p}{2T}-\coth\frac{E_{q-p}}{2T}\Bigr] 
\Bigl[\frac{1}{\omega+\frac{v^2p}{E_p}q+{\rm i}\delta}
-\frac{1}{\omega-\frac{v^2p}{E_p}q+{\rm i}\delta}\Bigr] \nonumber \\
&&+\mbox{regular terms}.
\end{eqnarray}
Thus it has a non-diffusive pole at $\omega=\pm (v^2p/E_p)q$
which can not be removed by finite temperature effects.
This pole gives a logarithmic singularity $\sim \log\omega$
of $1/T_1$ which signifies the ballistic spin transport.
The result is basically equivalent to that obtained for the free boson model
by Jolicoeur and Golinelli.\cite{jol}
This ballistic behavior is due to the cancellation 
between the damping effect and the vertex corrections.
Although we have demonstrated the cancellation only 
up to the lowest order in $1/N$,
we expect that the cancellation holds in all orders.
In this model the forward scattering is the dominant low-energy process.
Thus the quasiparticle damping does not give rise a diffusive behavior,
if one takes into account the vertex corrections which preserve
the current conservation law.
The above result is consistent with the Bethe ansatz solution
obtained in Sec. II.
Therefore we can conclude that the spin diffusion is absent 
in the 1D quantum non-linear sigma model itself.

\section{Coupling with Phonons}

In this section, we will discuss about the origin of
spin diffusion observed experimentally in the Haldane gap system,
${\rm AgVP_2S_6}$.
The important feature observed in this experiment is
that there exists the cutoff frequency $\omega_c$ 
which depends on temperatures like $\sim \exp(\Gamma/T)$ 
with $\Gamma$ a constant.\cite{taki}
At $\omega>\omega_c$, the spin diffusion is observed.
One possible mechanism of the spin diffusion 
may be a coupling with some other degrees of freedom, 
such as phonons, as discussed by Narozhny 
for the $s=1/2$ Heisenberg spin chain.\cite{naro}
Here we will investigate this possibility.
In order to treat the interaction between spin and phonon,
we invoke the $1/N$-expansion for the $O(N)$ sigma model.
We add the following term to the Hamiltonian which represents
the spin-phonon interaction,
\begin{equation}
H'=g'\int{\rm d}x(b^{\dagger}(x)+b(x))(\partial_x\vec{n})^2,
\label{phonon}
\end{equation}
where $b(x)$($b^{\dagger}$(x)) is an annihilation(creation) operator
for phonons, and $g'$ is a coupling constant.
For the case of $N=3$, this term comes from the continuum limit
of the spin-phonon interaction defined on lattices,
\begin{equation}
H'=J'\sum_i(b^{\dagger}_i+b_i)\vec{S}_i\cdot\vec{S}_{i+1}.
\end{equation}
We expect that the analysis for the $O(N)$ sigma model gives
a qualitative understanding for the effect of this term.
For $N \rightarrow \infty$, we can treat the spin-phonon
interaction eq.(\ref{phonon}) perturbatively.
The diagram for the lowest order self-energy correction is 
shown in Fig.2. 
We can easily evaluate this diagram
using the Green's functions for massive magnons (\ref{magnon}) and
free phonons given by,
\begin{equation}
D(p,\omega_n)=\frac{c^2p^2}{({\rm i}\omega_n)^2-c^2p^2},
\end{equation}
where $c$ is the velocity of phonons.
Then the imaginary part of the retarded self-energy reads,
\begin{equation}
{\rm Im}\Sigma^R(q=0,\omega)\sim 
\frac{\omega}{T\sinh^2(\frac{m^{*}}{2T})}\frac{2cm^2 g'^2}{c^4-v^4}
\equiv C(T)\omega,
\label{damp}
\end{equation}
for small $\omega$.
Here $m^{*}=m/\sqrt{1-(v^2/c^2)}$. 
Then, if $C(T)>E_q$ holds, the retarded Green's function 
of massive $N$-plet $\mbox{\boldmath $n$}$ has a diffusive pole,
\begin{equation}
G^R(q,\omega)=\frac{D_s}{{\rm i}\omega-D_s q^2-(D_sm^2/v^2)},
\label{diff}
\end{equation}
where $D_s=v^2/C(T)$.
In contrast to the case without phonons discussed in Sec. III,
this damping effect does not cancel with vertex corrections
of the spin-spin correlation function, eq.(\ref{spinc}),
since the spin current operator is not conserved.
Neglecting vertex corrections and using the diffusive propagator
(\ref{diff}), we evaluate the RPA type diagrams for 
the spin correlation function (\ref{spinc}).
The result is given by,
\begin{equation}
\chi(q,\omega)\sim \frac{Z}{{\rm i}\omega-D_s'q^2-\omega_c},
\label{spindiff}
\end{equation}
for small $q$ and $\omega$.
Here $D_s'$, $\omega_c$, and $Z$ have the same 
temperature-dependence as $D_s$. 
Thus we have a diffusive behavior of the spin-spin correlation 
function.
The Fourier transformation of eq.(\ref{spindiff})
with respect to $\omega$ leads to,
\begin{equation}
\int^{\infty+{\rm i}\varepsilon}_{-\infty+{\rm i}\varepsilon}
\frac{{\rm d}\omega}{2\pi}
\chi(q, \omega)e^{{\rm i}\omega t}=D_s'e^{-[D_s'q^2+\omega_c]\vert t\vert}.
\end{equation}
Therefore, $\omega_c$ gives a cutoff frequency.
For $\omega>\omega_c$ the diffusive behavior appears.
The important point is that $\omega_c$ depends on temperatures,
$\omega_c \sim T\sinh^2(m^{*}/2T)$, if we neglect the
temperature dependence of the mass gap $m$.
These results seem to give a qualitatively satisfying explanation for 
the spin diffusion and the temperature-dependent cutoff 
observed in ${\rm AgVP_2S_6}$.
Note that the presence of the temperature-dependent cutoff frequency is
quite due to the mass gap of the spin excitations.
The above argument is applicable only if the damping effect 
eq.(\ref{damp}) is large enough to give rise to a diffusive pole.
In this case, the perturbative approach may break down, and
the corrections due to higher order processes are not negligible,
though it is expected that the propagator has a diffusive form, 
eq.(\ref{diff}), for small momentum and energy.
Thus it may be difficult to fit our result to the experimental data
precisely.
However tuning the parameters, $v$, $c$, $g'$ properly, 
we obtain the temperature dependence of $\omega_c$ which is
qualitatively consistent with the experimental results.

\section{Summary and Discussions}

In this paper, we have shown that the spin transport of the 1D quantum
non-linear sigma model is ballistic even at finite temperatures, and
that the spin diffusion does not exist.
We have also argued a possible mechanism for the spin diffusion observed
experimentally in the Haldane gap system ${\rm AgVP_2S_6}$.
It has been shown that the coupling with phonons accounts for
the experimental results qualitatively.

Our study is based upon the non-linear sigma model which is believed to
be a low-energy effective theory for Haldane gap systems.
We can not answer the question whether the spin diffusion intrinsically
exists or not in the $s=1$ Heisenberg spin chain model.
According to de Alcantara Bonfim and Reiter,
the 1D classical Heisenberg spin model does not show the spin diffusion
even in the high temperature limit.\cite{alc}
The result for the classical spin system may be independent of 
whether the spin $s$ is an integer or a half-integer, and generally
quantum effects may suppress the thermal spin diffusion.
Thus it is unlikely that the spin diffusion appears in
the $s=1$ quantum Heisenberg spin chain itself.
We speculate that the spin diffusion observed in 
${\rm AgVP_2S_6}$ may be due to
a coupling with some other degrees of freedom.
However we need further investigations in order to reveal
whether the spin diffusion is possible or not in
the pure Heisenberg spin chains with an integer spin $s$.

\acknowledgments{}
The author would like to thank N. Kawakami for useful discussions
and critical reading of the manuscript. 
He is also grateful to P. B. Wiegmann and M. Takigawa
for invaluable discussions.
This work was partly supported by a Grant-in-Aid from the Ministry
of Education, Science, Sports and Culture, Japan.


\newpage

\begin{figure}
\centerline{\psfig{file=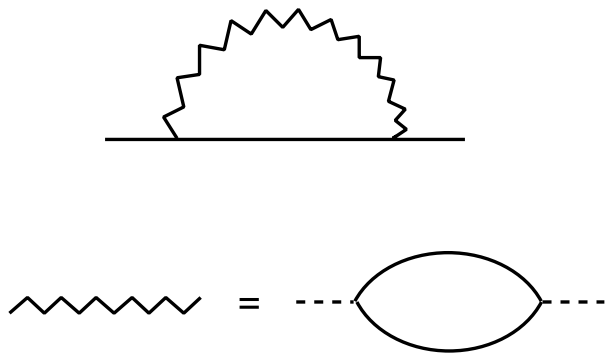,width=8cm}}
\vskip7mm
\caption{The diagram for the lowest order self-energy of
massive $N$-plet bosons. The solid line is the propagator for
the massive $N$-plet. The broken line represents the interaction
between the $N$-plet bosons.}
\end{figure}

\vskip10mm

\begin{figure}
\centerline{\psfig{file=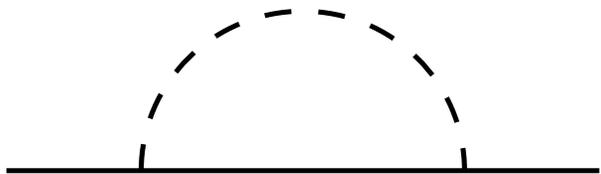,width=8cm}}
\vskip7mm
\caption{The diagram for the lowest order self-energy correction
due to phonons. The broken line represents the phonon propagator.}
\end{figure}

                                                                    
\end{document}